\documentclass[prb,twocolumn,superscriptaddress]{revtex4-1}
\usepackage{dcolumn}
\usepackage{bm}
\usepackage{graphicx}
\usepackage{amsmath}
\usepackage{amssymb}
\usepackage{ulem}
\usepackage[usenames]{color}

\begin{document}

%% materials
\newcommand{\yrs}{YbRh$_2$Si$_2$}
%% abbreviations
\newcommand{\ie}{{\it i.e.}}
\newcommand{\eg}{{\it e.g.}}
\newcommand{\Tkh}{$T_K^{high}$}
\newcommand{\Tk}{$T_K$}
\newcommand{\Tcoh}{$T_{coh}$}
\newcommand{\Rh}{$R_{\rm H}$}
\newcommand{\red}[1]{\textcolor{red}{#1}}

\title{Lifshitz transitions and quasiparticle de-renormalization
in YbRh$_2$Si$_2$}

\author{H. R. Naren}\altaffiliation{present address: Department
of Condensed Matter Physics, Weizmann Institute of Science, 76100
Rehovot, Israel} \affiliation{Max Planck Institute for Chemical
Physics of Solids, N\"othnitzer Str. 40, 01187 Dresden, Germany}
\author{S. Friedemann}
\affiliation{Cavendish Laboratory, University of Cambridge,
Cambridge CB3 OHE, UK}
\author{G. Zwicknagl}
\affiliation{Institut f\"ur Mathematische Physik, TU Braunschweig,
Mendelssohnstr.\ 3, 38106 Braunschweig, Germany}
\author{C. Krellner}\altaffiliation{present address:
Goethe-University Frankfurt, Institute of Physics, 60438
Frankfurt/Main, Germany} \affiliation{Max Planck Institute for
Chemical Physics of Solids, N\"othnitzer Str. 40, 01187 Dresden,
Germany}
\author{C. Geibel}
\affiliation{Max Planck Institute for Chemical Physics of Solids,
N\"othnitzer Str. 40, 01187 Dresden, Germany}
\author{F. Steglich}
\affiliation{Max Planck Institute for Chemical Physics of Solids,
N\"othnitzer Str. 40, 01187 Dresden, Germany}
\author{S. Wirth}
\affiliation{Max Planck Institute for Chemical Physics of Solids,
N\"othnitzer Str. 40, 01187 Dresden, Germany}

\date{\today}

\begin{abstract}
We study the effect of magnetic fields up to 15 T on the heavy
fermion state of \yrs\ via Hall effect and magnetoresistance
measurements down to 50 mK. Our data show anomalies at three
different characteristic fields. We compare our data to
renormalized band structure calculations through which we identify
Lifshitz transitions associated with the heavy fermion bands. The
Hall measurements indicate that the de-renormalization of the
quasiparticles, \ie\ the destruction of the local Kondo singlets,
occurs smoothly while the Lifshitz transitions occur within rather
confined regions of the magnetic field.
\end{abstract}
%\submitto{NJP}
\maketitle

\section{Introduction}
\yrs\ is a well-studied heavy fermion compound \cite{tro00} which
has an antiferromagnetic (AFM) ground state below $T_{\rm N} =$ 70
mK. The corresponding AFM transition can be suppressed to zero
temperature by the application of a small magnetic field $B_{\rm
N} =$ 60 mT (660 mT) perpendicular (parallel) to the
crystallographic $c$ direction of the tetragonal lattice
structure. Right at $B_{\rm N}$ indications for the existence of a
quantum critical point (QCP) have been observed \cite{geg02}. At
fields higher than $B_{\rm N}$, the system resides in a heavy
Fermi liquid state below a crossover temperature $T_{\rm FL}$.
There is an extended considerably large regime of non-Fermi liquid
(NFL) behavior fanning out above the QCP in the $B-T$ phase space.
This QCP is believed to be unconventional, in that it involves the
breakup of Kondo singlets at fields below a certain field $B^{\rm
Hall}(T)$ which coincides with the QCP at zero temperature, and
can be traced up to much higher temperatures \cite{pas04}.

The $J = 7/2$ multiplet state of the Yb$^{3+}$ ions is split into
four Kramer's doublets by the crystalline electric field (CEF)
within the tetragonal structure of \yrs. Kondo scattering over the
entire $J = 7/2$ multiplet gives a Kondo temperature estimate of
$T_K^{high} \approx$ 80 K from the minimum observed in thermopower
measurements \cite{koe08} (in our notation of a high Kondo
temperature, \Tkh, and a lower one, $T_K^{low}$, we rely on
Ref.~\cite{coq}). However, since the separation between the ground
state doublet and the first excited CEF state is large
\cite{sto06} ($\sim$200 K), the ground state doublet dominates the
Kondo scattering at low temperatures. The single-ion Kondo
temperature of the ground state doublet, $T_K^{low} \equiv T_K$,
has been estimated from entropy considerations via specific heat
measurements \cite{geg06} to \Tk\ $\approx 25$. Moreover,
thermopower measurements \cite{koe08} on magnetically diluted
Lu$_{1-x}$Yb$_x$Rh$_2$Si$_2$ yielded $T_K \approx$ 29 K for \yrs.
In general, there is a third energy scale, \Tcoh, which denotes
the temperature below which Kondo-lattice coherence sets in and
hybridized heavy fermion bands form. However, at least in the case
of \yrs, Scanning Tunneling Microscopy (STM) measurements
\cite{ernst11} provided evidence that the Kondo-lattice coherence
develops once the 4$f$ electrons have sufficiently condensed into
the CEF ground state, \ie\ that \Tk\ $\approx$ \Tcoh. Therefore,
we will use $T_K$ for this scale henceforth. One focus here is to
study the fate of the heavy quasiparticles in the Kondo systems
\yrs\ in high magnetic fields.

Applying a magnetic field to a heavy fermion system can cause
several effects: One is similar to increasing the temperature in
that the single-ion Kondo effect is increasingly weakened. Another
is Zeeman splitting which may become significant by inducing
Lifshitz transitions (LTs), \ie\ a band may get spin-split beyond
the Fermi energy $E_{\rm F}$, or the Fermi surface topology may
change drastically. Zeeman splitting is practically insignificant
in normal metals since the relevant energy scale $E_{\rm F}$ is of
the order of few eV which corresponds to magnetic fields of about
10$^4$ T. In heavy fermion metals, however, this scale is greatly
reduced due to the hybridization of conduction and localized $f$
electrons via the Kondo effect.

A further effect can be a metamagnetic transition which many heavy
fermion compounds undergo at a characteristic magnetic field
$\widehat{B}$ applied along the easy direction of magnetization. A
few examples are CeRu$_2$Si$_2$ ($\widehat{B} \approx$ 7.8 T,
Ref.\ \cite{flo95,flo02}), CeTiGe (12~T, Ref.\ \cite{dep12}),
UPt$_3$ (20 T, Ref.\ \cite{sug99}) and CeCu$_6$ (4 T, Ref.\
\cite{sag87}). The former three compounds show a sharp jump in the
field-dependent magnetization at $\widehat{B}$ whereas the latter
one only exhibits a kink at $\widehat{B}$. Moreover, in CeTiGe
there are indications \cite{dep12} for a first order phase
transition at $\widehat{B}$. The metamagnetic transition in
CeRu$_2$Si$_2$, which has extensively been investigated with
respect to this issue, was initially attributed to a destruction
of the Kondo effect resulting in the increase of magnetization
\cite{flo02}. The field scale $\widehat{B} =$ 7.8 T was believed
to correspond to the Kondo energy scale of $\sim$20 K in this
system beyond which the $f$ electrons were thought to become
localized. However, the metamagnetic transition was later argued
\cite{daou06} to result from a LT, a conclusion based on transport
measurements, model calculations and a re-interpretation of de
Haas-van Alphen (dHvA) results.

\begin{figure}[t]
\centering\includegraphics[width=8.4cm,clip]{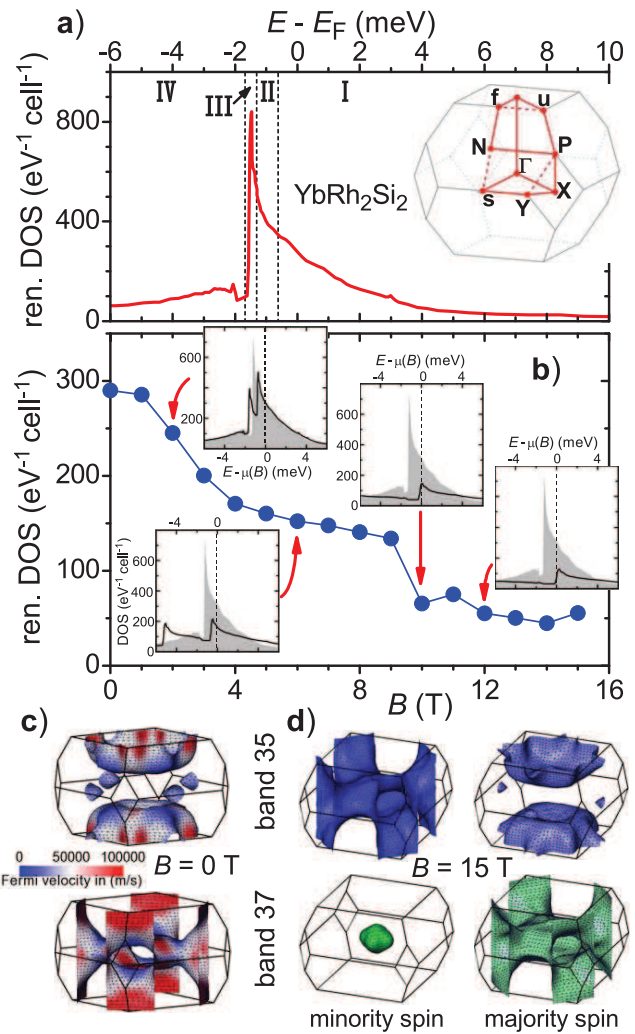}
\caption{Renormalized band structure calculations on \yrs. a)
Energies close to $E_{\rm F}$ displaying the van Hove singularity
and the division into four regions (marked I to IV) separated by
Lifshitz transitions (marked by dashed lines). Inset illustrates
the directions. b) Variation of the renormalized DOS with magnetic
field. Insets: DOS($E$) at different magnetic fields clearly
showing a Zeeman splitting of the van Hove singularity (for
comparison, the zero-field DOS is shown in grey in the background,
same scales are used for all insets). c) Calculated Fermi surfaces
for the two main bands 35 and 37. Colors indicate the Fermi
velocity. d) Fermi surfaces at 15 T for the same two bands and for
majority and minority spin direction.} \label{calc}
\end{figure}

A corresponding scale $\widehat{B} \approx$ 10 T for \yrs\ along
its crystallographic $ab$ plane (which is the easy plane of
magnetization) was estimated from a kink in magnetization
\cite{geg06}. Moreover, the quantities depending on the density of
states (DOS), like the magnetic susceptibility, the Sommerfeld
coefficient of the electronic specific heat $\gamma$, the
$A$-coefficient of the resistivity $\varrho$ (within Fermi liquid
theory $\varrho = \varrho_0 + AT^2$ where $\varrho_0$ is the
residual resistivity) and the linear magnetostriction coefficient
all decrease in a pronounced fashion around this field. The
decrease in DOS, {\it cf.} Fig.\ \ref{calc}(c), was interpreted as
a destruction of the heavy fermion state \cite{kne06} and
$\widehat{B} \approx$ 10 T could experimentally be related to the
Kondo energy scale via $k_{\rm B}T_{\rm K} \approx g\mu_{\rm
B}\widehat{B}$ ($k_{\rm B}$ and $\mu_{\rm B}$ are the Boltzmann
constant and the Bohr magneton, respectively; $g \sim$ 3.5 is the
$g$ factor \cite{sich09}). Another reasoning for the association
of $\widehat{B}$ and $T_K$ was based on the identical pressure
dependence of the two quantities \cite{geg06}.

A dHvA study of \yrs\ revealed \cite{rou08} a gradual reduction of
the dHvA frequency across $\widehat{B}$. This was interpreted in
terms of a LT, \ie, at $\widehat{B}$ one of the spin-split
components of a heavy band is shifted beyond the Fermi level.
Calculations based on static \cite{kus08} and dynamic mean field
theory \cite{bea08} endorsed the LT scenario to be responsible for
the anomaly at $\widehat{B}$. Another argument against an
alternative explanation, namely the destruction of the heavy
fermion state, would be the sizeable value of $\gamma$ of around
100 mJ/mol$\,$K$^2$ even beyond 10 T. This value is much larger
than the one reported \cite{fri10} for the local moment analogue
LuRh$_2$Si$_2$ (6.5 mJ/mol$\,$K$^2$).

In this work, we report on a high-resolution study of
magnetotransport (Hall effect and magnetoresistance) on
high-quality single crystals of \yrs\ and concentrate on high
magnetic fields (up to 15 T, in contrast to earlier reports
\cite{pas04,friPNAS} which focused on the QCP at small fields) in
order to shed light on these transitions. These measurements are
facilitated by renormalized band structure calculations to support
our assertions (for the ease of discussion we start off with
presenting these results first). While topological changes of the
Fermi surface may not necessarily reflect a significant change of
the DOS, they can create new open or closed orbits. Thus,
transport measurements can be a very sensitive tool to study such
changes.

\section{Renormalized Band Calculations with applied magnetic field}
The renormalized band calculations for \yrs\ have also been
extended to study the magnetic field evolution of the DOS
\cite{zwi11}. The influence of the magnetic field is accounted for
by field-dependent values for the centers-of-gravity
$\tilde{\epsilon}_{fm} (B)$ and effective widths
$\tilde{\Delta}_{fm} (B)$ of the renormalized $f$-bands which are
obtained from fits to the field-dependent quasiparticle DOS of the
single-impurity Anderson model \cite{hew04,hew06,hew07,pet06}. The
latter are calculated by means of the numerical renormalization
group (NRG). The isoenergy surfaces in zero field can be
correlated to the Fermi surfaces within a magnetic field in terms
of the position of the Fermi energy with respect to the van Hove
singularity of the minority spin DOS. These surfaces are expected
to be topologically similar without and with field even though the
height of the van Hove singularity itself decreases with field,
see insets to Fig.\ \ref{calc}(b).

Figure \ref{calc}(a) displays the zero-field DOS from which, in
particular, the partially developed hybridization gap and a van
Hove singularity can be recognized. These features are found below
$E_{\rm F}$ as expected for a hole (Yb-based) system. Four regions
can be identified within the investigated energy range within
which the isoenergy surfaces mainly keep their topology and which
are labeled I--IV in Fig.\ \ref{calc}(a) \cite{pfau0}. The
transitions between these regions are marked by LTs, dashed lines
in Fig.\ \ref{calc}(a).

With increasing field, the calculated DOS exhibits a progressive
reduction, with a marked jump at 10 T (see Fig.~\ref{calc}(b)).
These calculations indicated that the quasiparticle
de-renormalization, \ie\ the field-induced suppression of the
Kondo effect, takes over rather smoothly and hence, by itself
cannot create the anomaly at 10 T \cite{zwi11}. The field
evolution of the DOS as depicted in the insets of Fig.\
\ref{calc}(b) involves the Zeeman splitting of the zero-field DOS.
With increasing magnetic field the majority spin van Hove
singularity sweeps rapidly away from the Fermi level while the
minority spin van Hove singularity crosses $E_{\rm F}$ at around
10~T (see insets). In addition, the peak height of the van Hove
singularity reduces with increasing field owing to the
de-renormalization of the quasiparticles. Clearly, it takes the
renormalized band calculation including both the above-mentioned
effects as well as the quasiparticle interactions to reproduce a
field evolution of the DOS which conforms to the variation of the
Sommerfeld coefficient and thermopower \cite{pfau0}.

Fermi surfaces have been calculated for the two bands
predominantly contributing to the DOS. These two bands, band 35
and 37, give rise to the so-called `pillow' (upper picture in
Fig.\ \ref{calc}(c)) and `jungle-gym' (lower picture),
respectively. The color code in Fig.\ \ref{calc}(c) indicates the
Fermi velocity. Upon shifting the majority spin DOS($B$) to lower
energies by increasing the magnetic field the topology of its
Fermi surfaces remains largely unchanged, right column in Fig.\
\ref{calc}(d), since $E_{\rm F}$ stays within region I of the DOS.
In contrast, the minority spin Fermi surfaces strongly change when
the minority spin DOS($B$) moves up in energy and its
corresponding $E_{\rm F}$ travels through regions I--IV
\cite{pfau0}. Consequently, LTs are encountered in the minority
spin DOS($B$). The prominent one within band 35 is the formation
of a single, connected surface upon crossing from region II to
III, Fig.\ \ref{calc}(d). The isoenergy surface of band 37
undergoes a `neck-forming' LT in the crystallographic direction
$\Gamma$$\rightarrow$X between regions I and II (for directions
see inset to Fig.\ \ref{calc}(a)), followed by a `neck-disrupting'
LT at an angle between $\Gamma$$\rightarrow$X and
$\Gamma$$\rightarrow$s and a `pocket-disappearing' LT of the
pocket along X$\rightarrow$P$\rightarrow$u upon entering region
IV.

As is obvious from this insets to Fig.\ \ref{calc}(b), the
dominant peak of the minority spin van Hove singularity is shifted
beyond $E_{\rm F}$ at around 10 T. Since the width of this peak is
supposed \cite{cos02} to be the same as $T_K$, this again
indicates that the magnetic field scale of 10 T is indeed the
equivalent of $T_K$. From the magnitude of the shift of the
minority van Hove singularity, we can assign magnetic field values
to the transitions between the different regions: The
corresponding LTs are calculated to take place at $B_1 = 2.5 \pm
1$ T (region I to II), at $B_2 = 9 \pm 1$ T (region II to III),
and from region III to IV at $B_3 = 11 \pm 1$~T.

Depending on the extent of contribution of a specific band to the
DOS, the corresponding LTs are expected to cause changes in the
DOS($B$), Fig.\ \ref{calc}(b). A kink is seen at around $B_1$, a
drop at $B_2$ and a small maximum at $B_3$. However, the kink at
$B_1$ is, to some extent, already visible in the DOS($B$) just due
to the de-renormalization of the quasiparticles \cite{zwi11}.
Thus, the LT at $B_1$ seems to have a very minor effect on the
DOS. This can be understood \cite{zwi11} from the fact that the
dominant contribution to the zero-field DOS stems from the
`pillow' Fermi surface while it is the `jungle-gym' one that
undergoes a LT at $B_1$. The comparatively large jump at $B_2$ is
likely caused by both, the `pillow' and the `jungle-gym', sheets
being subject to LTs. In contrast, the faint feature at $B_3$ is
solely due to the LT of the erstwhile `jungle-gym' sheet whose
contribution appears to be more significant at high fields,
perhaps due to a reduced contribution from the erstwhile `pillow'
sheet. We note that the features at $B_1$ and $B_3$ were not
obvious in magnetization or heat capacity measurements
\cite{geg06}, but could be resolved in thermopower
\cite{pfau0,pourret}. In our magnetotransport measurements, we
clearly observe all these features as well as indications related
to the de-renormalization of quasiparticles.

\section{Experimental}
We performed {\it simultaneous} isothermal magnetoresistance (MR)
and Hall effect measurements down to $T \ge$ 50 mK and in magnetic
fields up to $B \le$ 15 T. To facilitate direct comparison,
current $j$ and $B$ were applied perpendicular to the
crystallographic $c$ direction for both, MR ($j \parallel B$) and
Hall measurements. Consequently, the Hall voltage $V_H$ was to be
measured along the $c$ direction. Since \yrs\ cleaves
perpendicular to the $c$ axis, we used two different crystals
(from the same batch, also same batch as in Ref.\ \cite{oes08})
with optimized geometries for the respective measurements. Note
that these are among the highest-quality crystals of \yrs\
(residual resistivity $\sim$$0.5 \cdot 10^{-8}\, \Omega$m). For
optimized sensitivity the sample for Hall measurements was thinned
down to 70 $\mu$m, and the signals were consecutively amplified by
low-temperature transformers, low-noise amplifiers and lock-in
amplifiers for both types of measurements. The actual Hall voltage
was taken as the antisymmetric component of the measured Hall
voltage under field reversal \cite{nair08}.

\section{Magnetoresistance}
\begin{figure}[t]
\centering\includegraphics[width=8.4cm,clip]{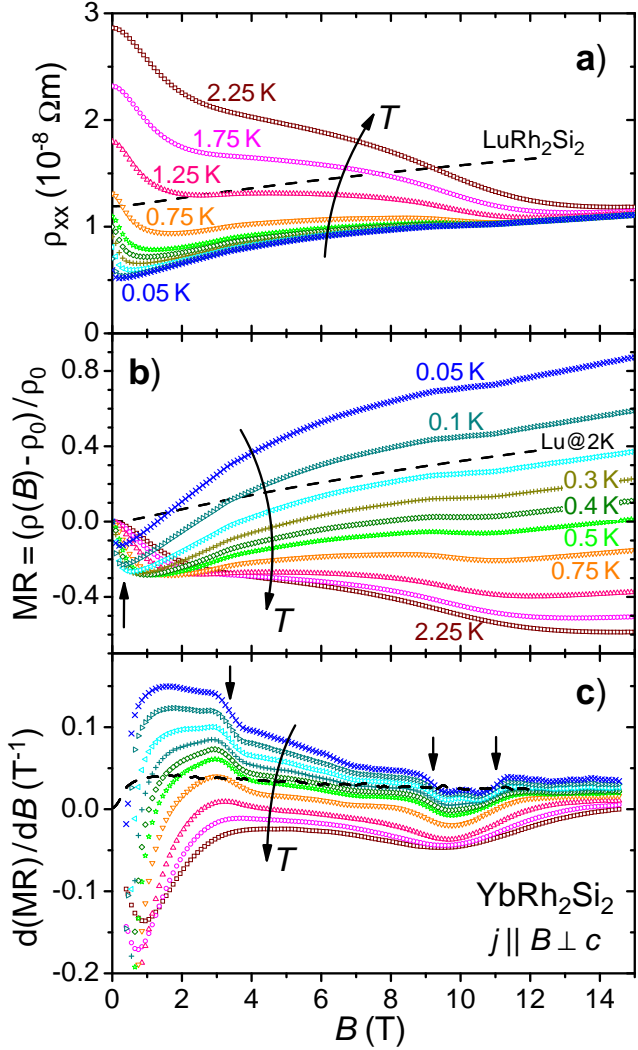}
\caption{a) Field dependence of resistivity $\varrho_{xx}(B)$, b)
of magnetoresistance and c) of its field derivative as functions
of magnetic field. All panels exhibit results at the same selected
temperatures, 0.05 K $\le T \le$ 2.25 K, and the same color code.
Black dashed lines: results of the nonmagnetic reference compound
LuRh$_2$Si$_2$ at $T =$ 2 K.} \label{MRall}
\end{figure}

Figure \ref{MRall} exhibits the field-dependent resistivity
$\varrho_{xx}(B)$, the magnetoresistance ${\rm MR} =
\frac{\varrho_{xx}(B) - \varrho_{xx}(B=0)}{\varrho_{xx}(B=0)}$ and
the field derivative of MR. The MR is positive at lowest
temperatures, as expected for the coherent state. At 50 mK and 15
T, the resistivity enhancement over the zero-field value is close
to 90\%. At very low fields, a step-like transition is visible
(marked by an upward arrow in Fig.\ \ref{MRall}(b)) at lowest
temperatures which gets smeared out quickly as temperature
increases. This has been reported to be a signature of the Fermi
surface reconstruction related to the unconventional QCP in this
compound \cite{friPNAS}. With increasing $T$, $\varrho_{xx}$ at
low fields increases due to progressive inelastic scattering of
conduction electrons. The negative MR at higher $T$ is then a
result of the magnetic-field suppression of the spin-flip
scattering.

At fields above 3 T, a small kink in the MR is observed that is
clearly reflected as a step in its derivative, marked by an arrow
in Fig.\ \ref{MRall}(c). A feature at this field scale has not
\begin{figure}[t]
\centering\includegraphics[width=8.0cm,clip]{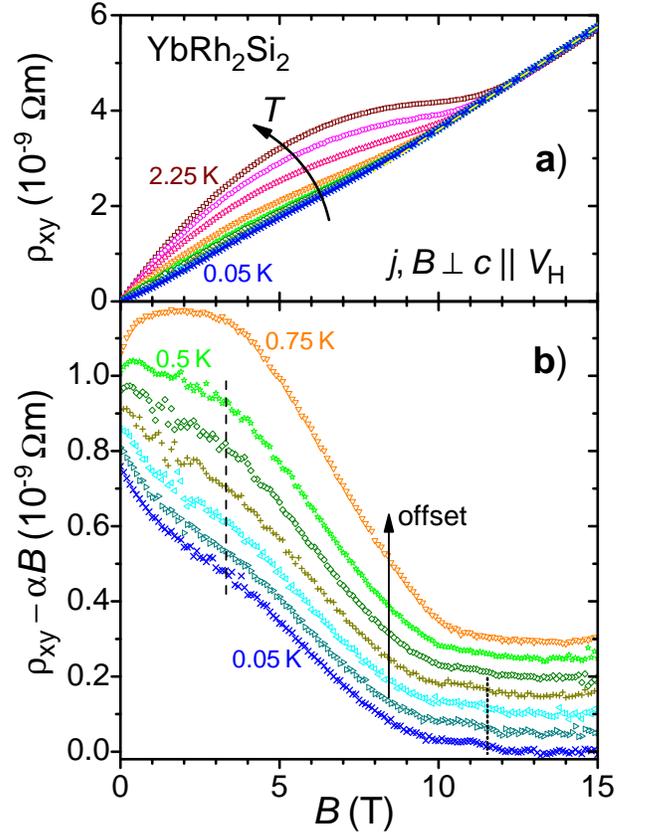}
\caption{Results of isothermal Hall effect measurements (same
temperatures and symbols as in Fig.\ \ref{MRall}). a) Field
dependence of Hall resistivity $\varrho_{xy}(B)$. The dashed line
indicates the linear high-field slope. b) Same low-$T$
$\varrho_{xy}(B)$ data as in a) but with the linear high-field
slope subtracted, $\varrho_{xy}(B) - \alpha \cdot B$. For further
clarity, curves (except at 50 mK) are offset by
0.5$\,\cdot\,$10$^{-10}\, \Omega$m. Dashed/dotted lines are guides
to the eye.} \label{rho_xy}
\end{figure}
been observed in previous measurements even though it should be
expected from the kink seen in the DOS($B$) \cite{zwi11}. The
anomaly observed at 10 T in Ref.\ \cite{geg06} appears as a double
kink in our MR data, again marked by arrows in Fig.\
\ref{MRall}(c). The two kinks are roughly at 9 T and 11 T and are
most sharply visible at lowest $T =$ 50 mK. Although these kinks
get smeared out at higher temperatures their positions in field
remain roughly the same. Thus, we indeed observe signatures of all
the three predicted LTs in our MR data.

At fields beyond 12 T, \ie\ beyond the high field anomaly, the MR
appears to becomes linear in field, at least at low temperature.
In fact, the linear regions in the low-temperature
$\varrho_{xx}(B)$-curves nicely overlap implying a temperature
independent high-field state. This point of view is further
supported by results obtained on the nonmagnetic reference
compound LuRh$_2$Si$_2$ which exhibits a featureless MR throughout
the measured field range (up to 12 T) with a slope very similar to
the high-field MR in \yrs. Such an increase of the MR is in line
with the existence of open orbits in the Fermi surface of \yrs\ at
higher fields, Fig.\ \ref{calc}(d).

\section{Hall measurements}
We now focus on the results of our Hall measurements presented in
Fig.\ \ref{rho_xy}. It has been shown that in \yrs\ at
temperatures below 1 K the anomalous contribution to the Hall
effect data is less than a few percent \cite{pas05}. In both
models considered in Ref.\ \cite{pas05}, the anomalous Hall
contribution is proportional to the magnetic susceptibility
$\chi$. Since $\chi(B)$ goes down for increasing fields $B$, the
anomalous contribution to the Hall effect is expected to continue
to be insignificant even at the high fields we have measured in.
In contrast, at higher temperatures the anomalous Hall
contribution becomes dominant. For example, the Hall resistivity
$\varrho_{xy}$ at $T =$ 2.25 K in Fig.\ \ref{rho_xy}(a) appears to
largely resemble the magnetization curve measured earlier
\cite{geg06}. We note here that, unfortunately, a comparison to
Hall measurements on the nonmagnetic reference compound
LuRh$_2$Si$_2$ was defied by the extremely small size of the
LuRh$_2$Si$_2$ single crystals.

The most intriguing result is the collapse of all measured curves
$\varrho_{xy}(T,B)$ at high fields into a single, linear-in-field
curve, \ie\ $\varrho_{xy}(T,B)$ appears to be independent of
temperature, see dashed line in Fig.\ \ref{rho_xy}(a). The field
value beyond which this collapse occurs increases with
temperature. Since the anomalous Hall contribution is small at
lowest temperature (see above), the temperature independence of
$\varrho_{xy}(T,B\gtrsim 12\, {\rm T})$ at high fields also
implies that the anomalous contribution becomes very small for
{\it all} measured temperatures at high fields. In turn, this
implies that the system at sufficiently high fields behaves
largely like an ordinary paramagnetic metal, even though it is
polarized. This view is corroborated by the fact that the
field-derived energy scale at which these ordinary metallic
properties occur corresponds to the energy scale $T_K^{low}$
relevant at the low temperatures investigated in the present
study.

At low temperature (below 0.5 K), the isothermal
$\varrho_{xy}(T,B)$ curves appear almost linear in $B$. However,
there are subtle changes of slope that become apparent if a
(large) linear ``background'' is subtracted. In Fig.\
\ref{rho_xy}(b), we plot $\varrho_{xy}(T,B) - \alpha \cdot B$,
where the constant $\alpha$ corresponds to the $T$-independent
high-field slope of $\approx\,$4.7$\,\cdot\,$10$^{-11}\;
\Omega$m/T. For clarity, an increasing offset (by
0.5$\,\cdot\,$10$^{-10}\, \Omega$m) was added to the
$\varrho_{xy}(T,B)$-curves above 50 mK. There is a clear
inflection point at around 3 T (marked by a vertical dashed line)
which corresponds to the inflection seen in the DOS($B$), Fig.\
\ref{calc}(b). This feature develops into a maximum at higher
temperatures, likely as a result of the additional anomalous
contribution to $\varrho_{xy}(T,B)$. Moreover, $\varrho_{xy}(T,B)$
at lower temperatures exhibits a step-like decrease at around 11 T
(dotted line cutting through low-$T$ curves only) which gets
smeared out at higher temperatures. This decrease, which seems to
appear at constant fields at different temperatures, is likely
related to the third LT at $B_3$.

\begin{figure}[t]
\centering\includegraphics[width=8.4cm,clip]{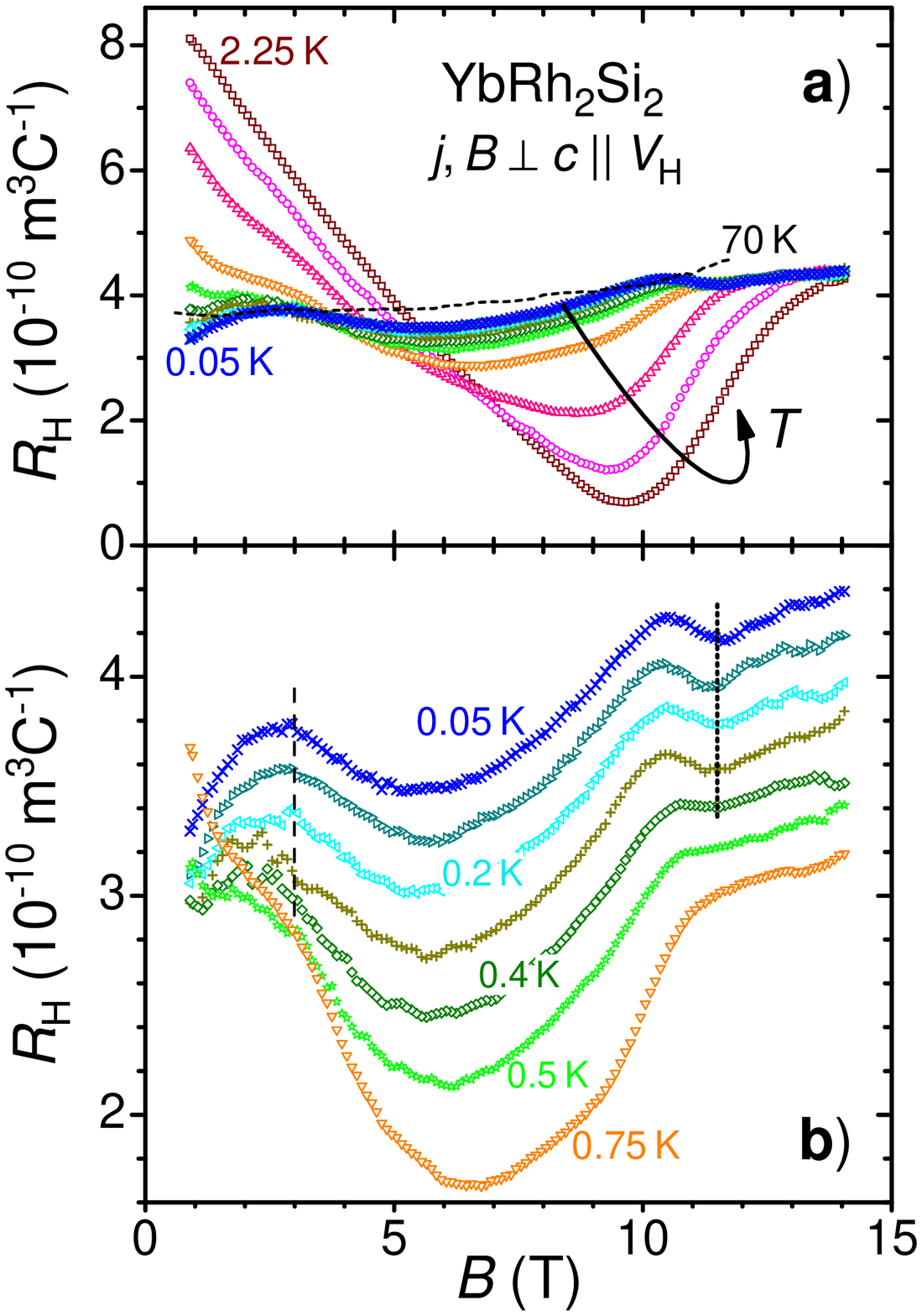}
\caption{a) Hall coefficient $R_{\rm H}$ (same temperatures and
symbols as in Figs.\ \ref{MRall} and \ref{rho_xy}). The dashed
line represents data at 70 K. b) Magnified view of the low-$T$
data. Curves at increasing $T$ are offset {\it downward} by
0.2$\,\cdot\,10^{-10}$ m$^3$C$^{-1}$ for clarity. At low $T$, the
normal Hall contribution dominates.} \label{Hall}
\end{figure}

\begin{figure}[t]
\centering\includegraphics[width=7.8cm,clip]{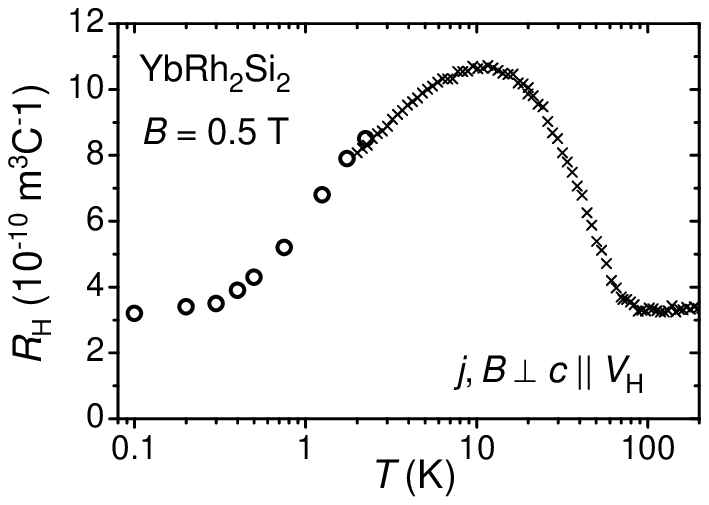}
\caption{Temperature dependence of the Hall coefficient \Rh\ at
0.5 T. Data marked by crosses ($\times$) were obtained on the {\it
same} sample but using a Magnetic Properties Measurement System
(MPMS).} \label{Hall-T}
\end{figure}

To gain insight into the evolution of the Fermi surface we now
consider the Hall coefficient $R_{\rm H} = \partial
\varrho_{xy}(T,B) / \partial B |_{T}$. The most prominent feature
in Fig.\ \ref{Hall}(a) is the minimum in \Rh\ at fields of roughly
9 T. This minimum strongly develops with increasing temperature
(above 0.5 K) and shifts its position towards higher field
indicating that it is {\it not} related to the DOS \cite{fert}.
Rather, it appears to be caused by fluctuations evolving upon
leaving the Fermi liquid regime with increasing $T$. Such behavior
is in line with a model \cite{col85} which describes the
temperature evolution of \Rh\ by skew scattering related to the
on-site Kondo effect, rather than coherent effects. This
temperature evolution of \Rh\ (as measured at 0.5 T) is presented
in Fig.\ \ref{Hall-T} and resembles the one obtained \cite{pas04}
for $B || c$. It confirms our conjecture above that \Rh\ is
dominated by the normal contribution, \ie\ it is related to the
DOS, only at lowest temperatures or at high fields $B \gtrsim$ 12
T. Inspecting the low-$T$ curves of \Rh, Fig.\ \ref{Hall}(b), an
anomalous Hall contribution appears to set in at $T =$ 0.75 K as
signaled by the dent observed around 9 T. We therefore concentrate
on the lowest measured temperatures in the following, Fig.\
\ref{Hall}(b).

At $T \le$ 0.2 K, a maximum in \Rh\ is observed around 3 T (dashed
line in Fig.\ \ref{Hall}(b)). In all likelihood this feature is
related to the LT at $B_1$, \ie\ the inflection in DOS($B$) and
above-mentioned neck formation along the $\Gamma$$\rightarrow$X
direction. Interestingly, among the three transitions, the one at
$B_1$ appears to have the most pronounced effect on \Rh\ resulting
in the corresponding maximum.

Upon increasing field there is a clear minimum in \Rh\ visible at
around 11.5 T and for $T \le$ 0.4 K, without apparent shift for
different temperatures (as indicated by the dotted line). One may
therefore speculate that this feature is related to the LT at
$B_3$. At this field, there is a maximum in DOS($B$), see Fig.\
\ref{calc}(b), along with severe changes in the topology of the
Fermi surfaces at these fields. The combination of these two
effects may account for the somewhat higher field values at which
the transitions are observed in \Rh\ compared to the calculations.
On the other hand, there is no clear feature seen in $R_{\rm H}$
in the field range around 9 T. As noted above ({\it cf.} Fig.\
\ref{calc}) there are two major bands at $E_{\rm F}$, both
undergoing LTs. We speculate that the transitions in these two
bands compensate each other such that the net change in $R_{\rm
H}$ is weak. We note here that thermopower measurements
\cite{pfau0} on a sample of the same batch and for identical
orientation also showed a maximum-minimum feature at fields around
11 T, but an additional, second maximum at around 9.5 T. While
this nicely corroborates our Hall data the occurrence of an
additional maximum also hints at the fact that electrical and
thermal transport measurements could be differently sensitive to
these phenomena.

At low $T$ an increasing background is visible in \Rh\ upon
increasing $B$, best seen in Fig.\ \ref{Hall}(b). This background
is even obvious in \Rh\ measured at 70 K (obtained on the same
sample but in a different measurement system limiting the absolute
quantitative comparison). Such an increase suggests a reduction in
the number of charge carriers (in the simplest model, \Rh\
$=-1/e\, n_{\rm eff}$ where $n_{\rm eff}$ is the effective charge
carrier concentration and $e$ is the charge of an electron). This
may be taken as another indication for the progressive
de-renormalization of quasiparticles at high magnetic fields, \ie\
of the on-site Kondo interaction. In other words, the
$f$-electrons seem to be gradually driven out of the Fermi volume
with increasing magnetic field. It nicely confirms the evolution
of the Kondo effect with decreasing temperature as discussed in
the introduction: the fact that the increase of \Rh\ with field is
still seen at 70 K, \ie\ below \Tkh\ but well above \Tk, clearly
points towards the single-ion nature of this effect.

Our measurements indicate a rather smooth
delocalization-localization transition at high fields. These
observations are in line with the already mentioned fact
\cite{geg06} that the Sommerfeld coefficient remains as large as
$\sim$100 mJ/mol$\,$K$^{2}$ beyond 10 T and is much larger than
the value of LuRh$_2$Si$_2$. In contrast, there is clear evidence
from renormalized bandstructure calculations that the observed
features at the different fields $B_1$, $B_2$ and $B_3$ could be
due to Lifshitz transitions which appear more abrupt in field.

A generic low-field Lifshitz transition was predicted \cite{ber12}
via DMFT calculations on the Kondo lattice model. Indeed, we do
observe such a transition in \yrs\ in our measurements.
%, it is interesting to note that measurements on CeTiGe
%\cite{dep12} and CeRu$_2$Si$_2$ \cite{pfau12} also appear to
%contain low-field features which could result from Lifshitz
%transitions.
In addition, Lifshitz transitions have been predicted to occur at
high fields in heavy fermion systems \cite{kus08,bea08}, at the
scale given \cite{ber12} by \Tk. However, we find two closely
spaced Lifshitz transition near 10 T. This could be due to a
slight difference in the coupling of the two bands to the magnetic
field.
%Field-dependent band structure plots could be very useful
%towards providing further clarity.

\section{Conclusion}
We have shown several Lifshitz transitions to occur in \yrs, by
severe changes of the Fermi surface topology of the dominating
bands vis-\`{a}-vis the shifting of the Zeeman-split Kondo
resonance through $E_{\rm F}$. While these transitions occur
rather abruptly, the de-renormalization of the quasiparticles
takes place comparatively smoothly. This phenomenology could be
generic among heavy fermion compounds, and magnetotransport seems
to be a useful tool in addressing such issues.\\[0.4cm]

We thank H. Pfau and M. Brando for insightful discussions. This
work is partly supported by the German Research Foundation through
DFG Forschergruppe 960.

\end{document}